\documentstyle[12pt,aaspp4]{article}

\def\eg{{\it e.g. }}
\def\etal{{\it et al. }}
\def\deg{\ifmmode^\circ\else$^\circ$\fi}
\def\ltsima{$\; \buildrel < \over \sim \;$}
\def\simlt{\lower.5ex\hbox{\ltsima}}
\def\gtsima{$\; \buildrel > \over \sim \;$}
\def\simgt{\lower.5ex\hbox{\gtsima}}

\begin{document}

\title{Diffuse Stellar Light at 100 kpc Scales in M87}

\author{Melinda L. Weil}
\affil{Department of Physics, Oxford University, Astrophysics Building,
       Keble Road, Oxford OX1 3RH, United Kingdom}

\author{Jonathan Bland-Hawthorn\altaffilmark{1} \& David F. Malin}
\affil{Anglo-Australian Observatory, P.O. Box 296, 2121 Epping, NSW,
       Australia}

\altaffiltext{1}{Visiting Fellow, Oxford University}
\begin{abstract}
In a new survey of nearby galaxies from stacked photographic images,
seemingly regular galaxies of several types show amorphous, often
asymmetrical features at very faint levels ($28 \hbox{mag
arcsec}^{-2}$).
In M87, a diffuse fan of stellar material extends 
along the projected SE (major) axis out to about 100 kpc. 

We suggest that accretion of a small spheroidal galaxy into a larger
potential is the most likely explanation for the diffuse structure.
The orbit is required to pass close to the center of the potential to
produce a fan which nearly aligns with the major axis and has a large
opening angle, as seen in M87.

Our simulations include a rigid primary potential with characteristics
similar to those derived for M87 and a populated secondary potential.
We investigate the structure of the dark matter at large galactic radii by
representing M87 with different potentials.  
The morphologies of the debris of
intruder spheres and disks of different masses and orbital
parameters limit the possible accretion scenarios.  
The total luminosity of the fan and the kinematics of debris 
in the center of the 
primary potential are analyzed and compared with substructure in M87.

The short lifetimes ($t_{fan} \simlt 5 \times 10^8$ years) of the
simulated diffuse fans 
and lack of observed shells indicates that either
we are seeing M87 at a `special time' during its evolution or that 
infall from small intruder galaxies is common.  
Our simulations indicate that several accretion events could be hidden in
galaxies.  For many orbits, intruder material is quickly spread out to
very low light levels.  Observations of the 
high specific frequency of globular clusters in M87 provide evidence that
the galaxy may experience frequent accretions of this type.

\end{abstract}

\section{Introduction}

A giant elliptical radio galaxy in the central regions of the nearby 
Virgo cluster, M87 (NGC 4486) exhibits properties that have 
stimulated numerous studies at all wavelengths.  Nonthermal emission is 
observed from the radio to the X-ray; the galaxy contains 
an optical synchrotron
jet, a small nuclear disk of ionized gas, and a broad distribution of 
X-ray emitting gas.  The luminous and compact nucleus is thought to 
surround a supermassive black hole which powers the galaxy's radio source. 
Although the Virgo cluster is considered an ideal host for a cooling flow,
searches for H\,I in absorption against the M87 radio source
fail to detect any spatially extended 
distribution (Dwarakanath, van Gorkom, \& Owen 1994).
The description of M87 as a classical E0 or E1 is
based on short exposure optical images; the optical isophotes, however, 
show marked eccentricity in deeper images.

The extremely uniform sensitivity and large sensitive area of 
photographic plates make them ideal detectors of faint light sources
extended over wide angles. This detectivity can be considerably improved
by `photographic amplification' (Malin 1978) and photographic `stacking'
or combination of several amplified derivatives (Malin 1988, 
Bland-Hawthorn, Shopbell, \& Malin 1993), extending
the detection to about 5 magnitudes below the night sky brightness. With
UK Schmidt plates at Siding Spring, this corresponds to a limiting (B)
surface brightness of about $28 \hbox{mag arcsec}^{-2}$. To produce
the image of M87 (Figure 1), five deep, IIIa-J (395-530nm passband) UK
Schmidt plates were combined.

Immediately apparent is the broad, diffuse and structureless `fan'
extending at least 15 arc minutes SE along the major axis of the galaxy.
Unfortunately, the image of a bright star (HD 109032, V=7.9, F0) lies in
this direction and probably adds a little to the extended structure.
However, a star of similar visual magnitude (HD 108614, V=7.95, K0) lies
about 40 arc minutes SW of M87 and has an inner halo hardly visible
against the night sky background, so the extended component of M87
evident in Figure 1 is real, though more symmetrical along the major axis
than the region contaminated by the star image would suggest.

An extension of M87 was first detected by Arp \& Bertola (1969) in
commendably deep exposures, which also showed continued isophotal
flattening at large radii.  These authors suggest that, while the
increasing ellipticity of the M87 isophotes is not expected if the
galaxy formed by collapse of a primeval cloud in which no angular
momentum transport occurs, the elongation could be the result of
ejection from the nucleus (Arp \& Bertola 1971).  The image in the
present paper is, however, deeper and emphasizes the asymmetry of M87 at 
large radii.  In it, M87 also appears to have a 
smaller, more regularly-shaped, caplike extension in the opposite direction 
to the SE fan.  

A concern is whether the extended fan is indeed stellar 
in origin. Other possibilities include optical line
emission or scattered light, both of which would
require an associated component of neutral gas.  Jura (1980) 
first suggested that the faint outer parts of spirals could 
arise as galaxian light scattered by the extended HI disk. 
His simple model predicted that it would be possible to 
distinguish between star light and scattered light at a 
visual magnitude of about 28-29 mag arcsec$^{-2}$. At these 
extremely faint levels, which are barely reachable with 
stacked photographic plates, it would be impossible to test 
for linearly polarized light with current technology. However, 
to explain the diffuse light as dust-scattered radiation from
M87 requires high column densities of gas ($\sim 10^{21}$ cm$^{-2}$)
with a total HI mass ($\sim 10^{10} M_{\odot}$) which would have
easily been detected in previous HI surveys. Therefore, it 
seems unlikely that the extended emission is associated with gas.
Nor has comparable structure been observed in either X-ray or radio maps 
(\eg Feigelson \etal 1987, Bohringer \etal 1995).

The present study was undertaken to determine whether production of 
large-scale diffuse structure, such as seen in this image of M87, 
and suppression of fine structure 
is possible in large-mass-ratio accretion 
events.  We explore the idea that the merger of a small intruder 
is responsible for the newly-discovered diffuse fan in M87.
Previous simulations of accretion focus on formation of fine structure
similar to that 
observed in a large fraction of elliptical galaxies (Malin \& Carter 1983). 
However, no shells have been observed in M87.  
Satellites which are accreted by larger 
galaxies with mass ratios $M_s/M_p = 0.01 - 0.1$ 
produce stellar shells and arcs with sharp edges 
(\eg Dupraz \& Combes 1986, Hernquist \& Quinn 1988, 1989). 
If gas is included in a disk intruder, it quickly segregates from the stellar
material to produce 
a clump or ring of gas in the center of the primary 
(Weil \& Hernquist 1993).  
In these previous models, roughly half of the stellar or gaseous particles  
are captured by the primary.  Little attention has been paid to the balance 
of the debris.

There is a wealth of evidence for accretion in ellipticals, including 
observations of multiple cores, shells, large quantities of dust and gas, 
and kinematic substructure including 
counter-rotating disks or rings.
Kinematic anomalies have been observed in M87; the mean velocity in the 
central regions is smaller than the inferred systemic velocity 
(Dressler \& Richstone 1990, van der Marel 1994).
Rotation curves of high spatial-resolution reveal 
a kinematically distinct 
subcomponent within $2^{\prime\prime} - 3^{\prime\prime}$ ($150 - 225$ pc for 
$D_{M87} = 15$ Mpc) of the nucleus (Jarvis \& Peletier 1991).  The 
blue-shifted annulus has a velocity of $40 \hbox{km s}^{-1}$ 
with respect to the
surrounding stars and covers an area of approximately $5^{\prime\prime}$.  
This subcomponent may be a dense core or  
disrupted material from a captured companion 
which has fallen to the center of M87.
In addition, the total number of globular clusters in M87 is $\approx 16000$, 
giving it a globular cluster specific frequency $S_N =14$ whereas the
typical specific frequency for normal ellipticals in dense regions is
$S_N = 4 -5$ (Harris 1991).  West \etal (1995) suggest that the excess 
population
of globular clusters in massive ellipticals in clusters may not be bound to 
the individual galaxy.  Instead, the globular clusters
occupy the cluster center as a whole, having been tidally stripped from
their parent galaxies during interactions or having originated in 
high-density regions of the cluster core itself.  
However, if accretion of small spheroids with $5-10$ globular clusters each 
is common over the lifetime of the galaxy, the excess population 
may be bound.  Dynamical friction considerations suggest that globular 
clusters, with many times the mass of a single 
star, could be stripped from the intruders and bound to 
the galaxy potential while much of the other stellar material escaped
or was dispersed.

In the following section, the methodology for constructing primary and
secondary potentials is presented, drawing on observations of M87 and of
small elliptical and disk galaxies.  $\S$ 3 presents the results of 
varying the initial conditions of the primary and intruder system 
and compares the structure of several models to M87 properties.  
In $\S$ 4 we discuss the results.

\section{Methodology}

The initial systems which we evolve comprise two galaxies: a massive 
elliptical which is intended to resemble M87 and a small intruder sphere or 
disk.  A large number of free parameters are available for exploration and
an attempt must be made to constrain the parameter space.   
Primary galaxy construction requires assumptions about the total mass, 
the shape of the potential, and relevant length scales.  We use recent 
observations of M87 to determine reasonable estimates for these quantities.
The intruding galaxy introduces a further number of free parameters into 
the simulations: galaxy type, mass, scale length, and the initial orbital 
position and velocity.

\subsection{Properties of M87}

At a distance of 15 Mpc, $1^{\prime\prime} = 75$ pc and $1^{\prime} =
4.5$ kpc in M87.  Figure 2 shows an image of M87 from the Digital Sky
Survey.  Visual band surface photometry of the stellar emission shows
that eccentricity increases from $\approx 10\%$ around $1^{\prime}$ to
$\approx 40\%$ around $3^{\prime}$ and remains this flattened beyond
$30^{\prime}$ (Carter \& Dixon 1978).  Axis ratios measured directly
from Figure 2 are $c/a = 1$ at $x = 3.4$ kpc and $c/a = 0.85$ at $x =12$ 
kpc, where $x$ is defined along the major axis.  From Figure 1, the
axis ratio at $x \approx 50$ kpc is $c/a \approx 0.6$.

The velocity dispersion of M87 is known to rise from about $300
\hbox{km s}^{-1}$ at 750 pc to $360 \hbox{km s}^{-1}$ in the center
(\eg Dressler \& Richstone 1990).  Long-slit spectrograph observations
out to 13 kpc show that the velocity dispersion declines from $\approx
375 \hbox{km s}^{-1}$ in the center to $\approx 325 \hbox{km s}^{-1}$
at 2.2 kpc, rises to $\approx 350 \hbox{km s}^{-1}$ at about 3.7 kpc, and
then declines to $\leq 300 \hbox{km s}^{-1}$ (Sembach \& Tonry 1996).
The zero point of the Sembach \& Tonry dispersion is $\approx 6\%$
higher than that of previous published studies; this is most likely
due to broadening associated with the large slit.  At large distances
from the center, there are velocity dispersion
anomalies which deviate from a smoothly declining profile.
These deviations may be caused by components with velocity
dispersions that differ from that of the underlying galaxy.
Although Sembach \& Tonry do not impute any significance to the deviations, 
the anomalies might be attributable to accreted material.
However, given the accuracy of the observations,
an approximately constant velocity dispersion from 7 -- 13
kpc cannot be ruled out.

Eccentricity in residual X-ray surface brightness isophotes 
along a NE--SW axis was discovered by 
Feigelson \etal (1987) after subtracting the axisymmetric component.   
They note that the gravitational potential, which is dominated by 
nonluminous mass at large distances from the center, is expected to 
affect the X-ray gas as well as the stellar component.
The ROSAT images of Bohringer \etal (1995) show that a few percent of 
the X-ray luminosity resides in a roughly linear structure extending 
across $10^{\prime}$.
A comparison with a radio map of M87 shows that
this thermal linear feature coincides with radio features in the outer halo 
and may be due to emission from cooled gas along the radio jets
rather than a response to the static gravitational potential.
Thus, there is no strong constraint for the shape of M87 at 100 kpc scales.

The enclosed mass at large radii has been calculated using X-ray data
and the kinematics of the globular cluster system.  Fabricant \& Gorenstein
(1983) find $1.2-1.9 \times 10^{13} M_{\odot}$ within $20^{\prime}$
(90 kpc), and $3.0-6.0 \times 10^{13} M_{\odot}$ within $60^{\prime}$
(270 kpc).  The ratio of mass to luminosity increases from $M/L=5 -15$ 
at $1^{\prime}$ to $M/L=50$ at $10^{\prime}$ and to nearly
$M/L=200$ at $20^{\prime}$, indicating that X-ray gas only accounts
for about 5\% of the dark matter in M87.  Stewart \etal (1984)
calculate the mass distribution using a number of models, all of which
converge to $M = 3\times 10^{13} M_{\odot}$ within $50^{\prime}$.
Merrit \& Tremblay (1993) use globular cluster velocities to constrain
models of the dark matter distribution and find 
$6^{+4}_{-1} \times 10^{12} M_{\odot}$ within 50 kpc, 
in agreement with the X-ray data.

Calculation of the core radius of M87 proceeds by assuming
that the dark matter core radius is similar to that of the globular
cluster distribution.  There is fairly good agreement among different
researchers, with values $R_c = 4.6 \pm 0.5$ kpc, $5.8$ kpc, and 
$6.6\pm 0.4$ kpc (McLaughlin 1995, Merritt \& Tremblay 1993, Lauer \&
Kormendy 1986, respectively).  
The large core of the globular cluster distribution likely 
represents the underlying dynamics of the formation epoch, as it appears 
unrelated to the luminous mass, and may be expected to reflect the dark 
matter distribution.  However, these results do not rule out a significantly
different value for the core radius; for example, Stewart \etal (1984) suggest
$R_c \approx 25$ kpc from models fit to the X-ray surface brightness 
distribution and spectroscopic measurements of the Fe lines.  
For comparison, the half-light radius of the stars in
M87 is 7.2 kpc and the core radius is an order
of magnitude smaller.

\subsection{Properties of the Primary Potential}

\subsubsection{Logarithmic potential} 

The axisymmetric logarithmic potential introduced by Binney (1981)
represents a galactic halo with a density that falls as $r^{-2}$ at large radii
with 
\begin{equation}
\Phi(R,z)= -{1 \over 2} v_0^2 log \left(R_c^2 + R^2 + {z^2 \over q^2} \right)
\end{equation}
where $R_c$ is the core radius, $v \rightarrow v_0$ at large r, and $q$ is 
the axial ratio of spheroidal 
equipotentials.  
This model was chosen because the axis ratios of the surface density contours 
decrease with radius as is observed for the luminous material of
M87.
The logarithmic potential models have a positive definite 
distribution function
for oblate galaxies with q in the range 
$0.707 \leq q < 1.0$ and for prolate galaxies with $1.0 < q \leq 1.08$;  
$q=1.0$ produces a spherical halo (\eg Evans 1993).
The cumulative mass distribution 
$M(r) = 4 \pi \int\limits_0^r \rho(r^{\prime})r^{\prime 2} dr$ 
is a function of $R_c$, $v_0$, and q
through the density
\begin{equation}
\rho(R,z)={v_0^2 \over 4 \pi G q^2} {\left(2q^2+1 \right)R_c^2+R^2+\left(
2-q^{-2}\right)z^2 \over \left( R_c^2 + R^2 +z^2q^{-2}\right)^2} .
\end{equation}  
For models with $q =$ 1.05, 0.975, 0.900 and 0.850 and a core radius 
$R_c = 6.0$ kpc, the requirement that the calculated masses resemble 
those measured from 
M87 X-ray data at large radii helps constrain $v_0$.  Taking the 
lower limit on the mass of M87 of 
$M(R=120\hbox{ kpc}) \approx 5 \times 10^{12} M_{\odot}$ and 
$M(R=240\hbox{ kpc}) \approx 10^{13} M_{\odot}$
requires $v_0 \approx 400 \hbox{km s}^{-1}$. 
$M(R=240\hbox{ kpc}) \approx 3 \times 10^{13} M_{\odot}$ implies 
$v_0 \approx 650 \hbox{km s}^{-1}$.

The projected surface density is calculated from 
\begin{equation}
\Sigma(x,y)={v_0^2 q \over 4G} {x^2 + y^2 + R_c 
\left(1+\cos^2i+q^2\sin^2i \right) \over \left[ \left( x^2 +R_c^2 \right) 
\left(\cos^2i+q^2\sin^2i \right) + y^2 \right]^{3/2}} 
\end{equation}
for an inclination angle $i=90\deg$.
The axis ratios for a $q=0.850$ logarithmic potential at 
$x =$ 3.5, 12, and 50 kpc are $c/a =$ 0.80,
0.70, and 0.65, respectively.  For comparison, the axis ratios at the 
same $x$ for $q=0.900$ are $c/a =$ 0.85, 0.80, and 0.75.

\subsubsection{Hernquist potential} 

For comparison, models are also generated from axisymmetric 
Hernquist potentials (Hernquist 1990).
These models provide good fits to the $R^{1/4}$ law 
light distribution in elliptical galaxies.
The local mass density is found from 
\begin{equation}
\rho = {M \over 2 \pi abc}{1 \over \mu \left(1+\mu\right)^3}
\end{equation}
where 
\begin{equation}
\mu^2 = {x^2 \over a^2}+{y^2 \over b^2}+{z^2 \over c^2} .
\end{equation}
The potential for axisymmetric cases is not analytic,
\begin{equation}
\Phi = -{GM \over 2} \int\limits_0^{\infty} {du \over \Delta(u) (1+m(u))^2}
\end{equation}
where $\Delta(u)=((a^2+u)(b^2+u)(c^2+u))^{1/2}$ and
\begin{equation}
m(u)= {x^2 \over a^2+u} + {y^2 \over b^2+u} + {z^2 \over c^2+u} .
\end{equation}
The enclosed mass is 
\begin{equation}
M(\mu)=M_{total} {\mu^2 \over (1+\mu)^2} .
\end{equation}
We tested models with spherical shapes: $a=b=c=1$; oblate shapes: 
$a=b=1$, $c=0.6$ and $c=0.8$; and
prolate shapes: $a=1$, $b=c=0.6$ and $b=c=0.8$.
The axis ratios are constant and allow a comparison with the effects of the 
varying axis ratios of the logarithmic potential.

Whereas the logarithmic potential has a large core defined by the
value $R_c$, the Hernquist potential is cuspy.  The top frame of Figure 3
shows the density profile for the two potentials with their masses
normalized to $M_{total} \approx 3 \times 10^{13} M_{\odot}$ 
at large radii.  For the logarithmic potential, the thin line
represents $R_c=6$ kpc and the thick line $R_c=25$ kpc, and the half-mass
radius is approximately 120 kpc.  For the Hernquist potential, the spherical 
model and the axisymmetric model with $c/a=0.8$
have similar density profiles in this representation.
The thin and thick lines represent models with $a=6$ kpc and $a=25$ kpc, 
respectively, where the half-mass 
radius is found from $r_{1/2} = (1+\sqrt{2})a$ in spherical potentials.

\subsubsection {Plummer and perfect potentials} 

For further comparison, models with Plummer and perfect potentials
are constructed.  Both of these models have finite cores.  The density of
the spherical Plummer model is 
\begin{equation}
\rho = {3M \over {4 \pi a^3 (1 + r^2/a^2)^{5/2}}}
\end{equation}
and 
that of the perfect model is 
\begin{equation}
\rho = {Ma \over {\pi^2(a+r)^2}} .  
\end{equation}
These models
are shown in the bottom frame of Figure 3, where the thin and thick lines
represent the different values for $a$ as above.  In the same panel, the
density profiles of the axisymmetric perfect model with an axis ratio $c/a=0.8$
and the isochrone model with a linear scale-length of 6 kpc are
shown for comparison.

\subsection{Properties of the Secondary Potential}

In the attempt to account for the broad, diffuse 
fan in M87 and the apparent lack of sharp-edged fine structure, we focus 
on encounters that might be expected to suppress strong shell formation.
Debris from the disruption of disk companions tends to show sharp-edged 
features even in the escaping stellar tails (\eg Weil \& Hernquist 1993, 
Figure 2).  In addition, little or no H I appears to exist in M87.  
Observations of gas and dust in ellipticals indicate that they are 
accreted in mergers.  The H I content of ellipticals is usually contained in 
disks or rings in 
a nuclear condensation or outside the optical boundaries of the galaxy; 
these structures are often rotationally supported, with larger specific 
angular momenta than the stars.
Even small disks contain enough neutral gas 
($M_{gas} \sim 10^9 M_{\odot}$) to pollute gas-poor early-type galaxies.  
Both of 
these considerations suggest that the galaxy responsible for M87's diffuse
fan was probably a spheroid.

A Plummer model with 
\begin{equation}
\Phi= -{ GM_s \over (r^2 + a_s^2)^{1/2}}
\end{equation} is used to represent small spheroids.  10,000 particles 
are randomly distributed in a sphere of radius $3 - 5 a_s$.  Because spheroids 
are supported by velocity dispersion, the initial velocities are set to 
approximate virial equilibrium.  Masses of 
$2 \times 10^{9} - 2 \times 10^{10} M_{\odot}$ 
and scale lengths of $300 - 1200$ pc are considered.

In order to test the influence of intruder type, we also simulate 
an infalling disk.
The disk has particles which are
distributed according to an exponential surface density profile.  The
total mass is $10^{-4}$ times the primary mass, the potential
scale length is 2 kpc, and the density scale length is 1.5 kpc.  The
disk stars are rotationally supported with velocities initialized
to provide centripetal equilibrium in the secondary potential.  As the
effects of velocity dispersion are ignored, stars follow circular
orbits in the small galaxy until it is tidally distorted by the primary.

\subsection{Evolution Code}

In the evolution of the systems, the primary galaxy is treated as a rigid 
potential whereas the small galaxy is rigid but populated.  
The code uses a variant of 
a restricted 3-body method which solves the 
two-body problem of the primary and secondary potentials,
based on the initial orbital parameters provided for each encounter.
For some models, as with the axisymmetric Hernquist potential whose 
acceleration is not 
analytic, the code is joined with the self-consistent field 
method (Hernquist \& Ostriker 1992) in which the primary 
density and potential 
are expanded in a set of basis functions in order to solve Poisson's equation.
A leapfrog integrator is used to compute the positions and velocities
of the center of mass of the galaxy and intruder at equal timesteps.
Then, the positions and velocities of particles are updated by
interpolating the tabulated phase-space coordinates for the two-body
system to calculate the acceleration.
The small and primary galaxies are initially separated by a sufficiently
large distance that tidal effects are negligible.  

The response of the secondary to the disrupting influence of the primary 
must be considered.  In previous simulations of large-mass-ratio accretions,
a `disruption radius' is chosen (\eg Hernquist \& Quinn 1988, 1989).
When the intruder reaches a distance $r_{disrupt}$ from the
primary, the potential
used to represent the small galaxy vanishes and the particles evolve 
in the solitary
gravitational field of the primary.
A concern in these simulations is whether the choice of disruption
radius strongly affects the subsequent evolution of the particles.
For previous simulations of shell formation Heisler and White (1990)
noted that, while the populations of shells depend on the details of
disruption of the companion, positions of shells depend only on the
potential of the primary.  However, for any one orbit, the opening
angle of the `fan' is determined by the differences in the tidal force
felt by the distribution of intruding particles.
The different distances of closest approach to the primary potential center 
will produce different velocities perpendicular to the original orbit for
each particle, and the tidal encounter will also induce different velocities
along the original orbit.  Thus, the opening angle will depend on the
range of the ratio of these two velocities.

A more self-consistent treatment, which takes dynamical friction into
account, is necessary to achieve the correct binding energy
distribution.  Here, we compare several methods: 1) specifying an
instantaneous disruption radius, 2) allowing the secondary potential
to survive the encounter with the large galaxy even when the
constituent particles of the intruder are torn away, and 3) slowly
disrupting the secondary potential by reducing its mass as particles
become unbound.  For the first method, a two-body integration is
used to determine the distance of closest approach which then becomes
$r_{disrupt}$.

\section{Results}

Three main ideas motivated the evolution of these models.  
First, we determine whether diffuse structure on scales of 
100 kpc is produced during the merger of a low mass intruder with a large 
elliptical galaxy.  Following that, we determine whether our 
models can constrain the shape of M87's potential; that is, is it oblate or 
prolate and does the observed variance of the axis ratios extend into the
non-luminous matter?  Finally, we attempt to use the persistence of the 
diffuse fan to estimate the time since the most recent accretion event and
to place limits on the number of recent accretions.  
Several initial systems were simulated and evolved in order to test the 
effects of such characteristics as 1) the shape of the primary potential,
2) the initial orbit, 3) different intruder masses and scale lengths, 
4) spheres versus disks, and 5) the disruption radius.  
These effects are discussed below and specific cases which produce structure
resembling that discovered in M87 are presented in detail.

The velocity dispersion of the Virgo cluster is $\sigma = 632 \pm 41
\hbox{ km s}^{-1}$ (Fadda \etal 1996).  Thus, we might expect the nearby
companions of M87 to have large relative velocities.  This would lead
to high-velocity, plunging mergers if the orbit of a small galaxy passed
close to the center of M87.  It is likely that M87 has
previously devoured nearby companions which might spiral slowly 
into its gravity well and that its present
source of merger material is small galaxies that have been diverted
from their original orbits by its deep potential well.  Therefore, in
most of the simulations the initial separation distance between the
two galaxies is 120 kpc and the intruder has zero initial velocity
relative to M87.  At the distance of closest approach, the relative velocity 
is very large, $v \simgt 1000 \hbox{ km s}^{-1}$.

The primary + intruder systems are evolved for at least 
several times $10^8$ years.  
During the evolution, the small galaxy plunges into the primary 
potential, passing near to its center, and is disrupted.  Generally, nearly
all of the stars form a spray of material at 100 kpc scales,
whose opening angle is dependent on the
initial orbital parameters and the shape of the primary potential.  
Some of this material eventually escapes 
the gravitational well of the primary, becoming too diffuse for detection at 
the present limits on timescales of the order of a few hundred million years.
The balance of the stars form 
finer structure nearer the center of the galaxy, often forming shells at 
late times.  However, previous to forming shells, on their second passage 
through the center of the galaxy, the intruding particles often form another
diffuse extension, opposite to the fan, which sometimes 
appears similar to that in M87.

\subsection{Varying Initial Conditions}

We first examined the effect on the evolution of diffuse structure of
discontinuing the secondary potential's attraction for its test
particles.  Several identical systems were evolved with an
$r_{disrupt}$ defined by the distance of closest approach, with no
disruption radii, or with slow disruption as the secondary mass is
decreased by the mass of its unbound particles.  For some potentials
and orbital parameters, there is little effect regardless of which
method of disruption is used.  When the secondary potential remains
intact, particles continue to be accelerated along the orbit of the
intruder long after it has been destroyed.  Slight enhancements of
density along the orbit are noticed at late times; however, models
with no disruption and slow disruption produce fans of the same size
and shape.  Because some models only produced healthy fans when the
secondary was turned off at $r_{disrupt}$, only models
with the other two methods of disruption are considered.

Tests of different initial positions, whose orbits have 
different distances of closest approach to the primary center, 
indicate that only orbits which pass near the center produce diffuse fans
which resemble that of M87.  However, the slight offset of the fan in M87 
from the major axis suggest that the encounter was not strictly radial.
The distances of closest approach vary from some 10s of parsecs to 
roughly 1 kpc for the simulations discussed below; however, there is no
correlation between the opening angles of the fan and the distances of
closest approach.  In all the models, the intruder passes close enough to
the center of the primary that the subsequent spread of the particles 
is not primarily dependent on the differences in closest approach.
Other tests with a disk intruder are shown to produce fans which
have sharp edges, unlike that in M87.  
Of the several models we evolved, a selection which encourages 
comparisons between
different initial conditions have their properties 
summarized in Table 1.  In the second and third columns, the primary and
secondary masses are listed; the fourth column contains the scale length of the
secondary.  The fifth and sixth columns contain the initial positions.
In the seventh column, a description of the primary is given.

To test the effect of different distributions of mass, primary potentials
with large and small cores and with cusps were constructed.  For the
logarithmic
potential, models with different degrees of flattening, 
$q =$ 1.0, 0.90, and 0.85, were considered.  
The mass of these galaxies is $M_p = 1 - 3 \times 10^{13} M_{\odot}$ 
at 240 kpc and the core radius is 6 kpc, similar to the values for
M87.  Intruders with masses $M_s = 10^9 - 10^{10} M_{\odot}$ and scale lengths 
$a_s = 0.3 - 1.2$ kpc are released from the position $x_i = 119$ kpc, 
$y_i=0$ kpc, and $z_i=-13.7$ kpc, a distance of 120 kpc from the center.  
There were no important differences in the evolution of models with no 
disruption and models which were disrupted slowly.  

The top panels of Figure 4 show three times 
during the evolution of Run L1, which has
a spherical logarithmic primary with $q=1.0$ and 
a secondary with mass $M_s = 1.2 \times 10^{10} M_{\odot}$ and scale length
$a_s=1200$ pc.  
The middle three panels show Run L2, the same as Run L1 except 
with a flattened primary of $q=0.85$.
The length of each panel is 360 kpc and times in 
simulation units, of which one equals $1.2 \times 10^6$ years,
are shown at the top of each. 
The ellipses show the shape of the primary surface density at 
$x =$ 50 and 120 kpc in the $x - z$ plane; as do M87's, the isophotes flatten 
with increasing distance from the center.
The accelerations on the intruder as it passes near the center of
the logarithmic potential are not sufficient to cause the particles to spray
into a fan as in M87; instead a very thin extension is produced.  
Similar results are found for the other tests of the logarithmic potential, 
including a prolate model with $q=1.05$;
no wide fans were produced.
However, the debris does not evince any fine structure.  These high-mass 
ratio accretion events do not form shells, which is encouraging for our 
hypothesis that such events can account for diffuse structure in galaxies
that have no shells.

In order to test the effect of the large core of the logarithmic
potential on the subsequent distribution of debris from the intruder,
a similar model with $q=0.85$ but $R_c=2.0$ kpc was evolved.  The
bottom panels of Figure 4 show three times during the evolution of
this model.  A fan with a wider opening angle is produced; however, it
is not comparable in size to that of M87.

In Figure 5, the results for three spherical models with
greater central concentrations of mass are shown.  The intruder
particles are shown after they have passed through the center of the 
potential and have spread significantly. 
The model in the top row, Run H1, has a Hernquist primary potential. 
The model in the middle row, Run P1, has a Plummer primary, 
that in the bottom row, Run P2, a perfect
primary.  Each has $M_p \approx 3 \times 10^{13} M_{\odot}$ at 240
kpc, a scale length of 6 kpc, and an intruder like those described for
the models in Figure 4.  The Plummer and perfect models with finite cores 
produce diffuse fans, but ones with 
smaller opening angles than that produced by the cuspy Hernquist model.
Nearly all of the intruder debris forms a
large, diffuse fan with an opening angle similar to that of M87 and
which travels along a narrow cone at a 
small angle from the long axis of the galaxy, as is seen in M87.
The finer structure, such as shells, that is not observed
in M87 is suppressed.

The effect of flattening and the shape of the primary were further 
explored for systems whose primary galaxies 
were constructed from a Hernquist potential.  Figure 6 shows
the evolution of Run H2
in the $x - z$ plane for a primary and intruder system similar to Run H1 
but with an oblate primary with $c/a=0.8$.
The length of each panel is again 360 kpc and times in 
simulation units are shown at the top of each panel.  
The ellipses show the shape of the primary surface density at 
$x =$ 50 and 120 kpc. The top row shows the
initial position in the $x - z$, $x - y$, and $y - z$ projections.  
Subsequent panels show the evolution in the $x - z$ plane.  
The intruder is destroyed by the tidal
forces acting on it.
The top panels of Figure 7 shows three times for a model, Run H3, 
exactly similar to Run H2 but with $c/a=0.6$, and the bottom panels are 
for Run H4, a prolate primary with $c/a=0.8$.
It is apparent that the flatter primary restricts the opening angle 
of the diffuse fan; the velocities perpendicular to the orbit are somewhat 
inhibited at large distances away from the center of the galaxy.
Although the accelerations of axisymmetric Hernquist potentials are not 
analytic, equations 6 and 7 imply that ${da_x \over da_z}$ is proportional to 
factors of $x/z$ and $c/a$.  Comparing Run H3 with Runs H1 and H2 indicates
that, after the intruder passes through the center of the
galaxy, the compression of the potential along the minor axis ensures 
reduction of the spread of accelerations in the z-direction.
That a fan comparable to that of Run H2 is produced by the prolate primary
of Run H4 suggests that it will be difficult to place strong limits on the 
shape of M87.

The mass of the intruder also has an effect on the shape of the 
disrupted material.  In previous models, the small galaxy had mass
$M_s = 1.2 \times 10^{10} M_{\odot}$ and scale length
$a_s=1.2$ kpc.  The top panels of Figure 8 show results for Run H5 in which a 
smaller intruder has $M_s=1.2 \times 10^{9} M_{\odot}$ and scale length
$a_s=300$ pc.  
The middle panels show results for Run H6 in which the intruder has
$M_s= 6.0 \times 10^{9} M_{\odot}$ and scale length $a_s=600$ pc. 
The bottom panels of Figure 8 show the fan produced by Run H7, 
which has a much
more massive intruder with $M_s= 6.0\times 10^{10} M_{\odot}$ and scale length
$a_s=2.0$ kpc.
A comparison with Run H2, in Figure 6, shows that one effect of increasing
intruder mass is to increase the opening angle of the diffuse fan.
By time 105, $1.3 \times 10^8$ years after the initial time, one half of the
mass is no longer bound to the secondary galaxy in Runs H5, H6, and H7.
At the distance of closest approach, at approximately time 106, the intruder
mass has been reduced by a factor of ten.  
Figure 9 is a velocity histogram for the particles in Runs H5, H2, and H7, 
in which the top frames are for times when most of the mass is still 
bound to the intruder and the bottom frames are for subsequent 
times, when the most of the mass is freely circulating in the potential of the
primary.  
The x-velocity histograms are shaded with horizontal lines; 
the z-velocity ones are shaded with lines at an angle.
It is apparent, in the top frames, that $v_z$ is uniformly small, as expected.
In the bottom frames from left to right, after the slow disruption of the 
intruder, the spread in $v_z$ increases as secondary galaxy mass increases.

Figure 10 shows the evolution of Run H8 in which the scale length of the
primary galaxy is $a_p=25$ kpc.  When the half-mass radius of the galaxy
representing M87 is about 60 kpc, the mass distribution is markedly different,
as shown in Figure 3; however, a fan, though not one as wide as that of Run H2,
is formed.
In Figure 11, an intruder with $M_s=10^{10} M_{\odot}$ and $a_s=1.2$ kpc 
is used, but the mass of the primary is
$M_p = 5 \times 10^{12} M_{\odot}$ at large radii.  
The top three panels are for a 
logarithmic primary, the middle ones are for an oblate Hernquist model with
$c/a=0.8$, and the bottom ones are for the same Hernquist model but with
the secondary initially positioned 60 kpc from the center of the primary, rather
than the fiducial 120 kpc; Runs L3, H9, and H10, respectively.  
The sizes of the diffuse fans show that reducing
the mass of the primary allows production of fans with larger opening angles.
In Run H10, the reduced velocity at closest approach restricts the x-extent and
z-extent of the fan.

\subsection{Evolution at Late Times}

In Figure 6, at time 280,
the beginning of shell formation is apparent for Run H2 
as the tidally ruptured particles oscillate back and forth
through the primary.
The amount of mass in the diffuse fan decreases with time: while 60\% of the
intruder particles are in the fan at time 240, only about 30\% are at time 280.
By time 220, an enhanced diffuse extension is produced opposite to the fan,
very similar to the one seen in the image of M87.

Figure 12 shows Runs H1 (top), H2 (middle), and H5 (bottom) at lates times 
in their evolution.  In Run H2, by time 340 the fan only contains 
20\% of the intruder mass and
weak, but distinguishable, shell formation is visible.
For Run H1, nearly 77\% of the intruder mass constitutes the fan at time 260;
by time 340, only 32\% remains in the fan.
While none of these models show very sharp-edged shells, 
the projected density is enhanced at the turning points of the orbits,
where particle velocities are at a minimum.  As shown in previous simulations 
of shell formation, the shells appear to propagate outwards as less 
tightly-bound particles reach their turning points.
No shells have been observed in M87; this fact and information derived from
the existence or lack of other substructure may contribute to
estimates of the age of the proposed accretion event in M87.

In no case, in all the models, does a strong fan appear for more 
than a few hundred million years.  In the next section, the models which
produce fans comparable to that of M87 are more severely constrained; we
determine which fans would be visible with the techniques available to us
and which produce approximately the amount of light estimated for the
M87 diffuse fan.

\subsection{Spatial and Kinematic Structure}

The total luminosity in the M87 fan is estimated using the Johnson B band
with a zero point of
$6.61 \times 10^{-9} \hbox{ erg cm}^{-2}\hbox{ s}^{-1}\hbox{ \AA}^{-1}$ and
a band width of 980\AA.  If the distribution of luminosity at 
$28 \hbox{ mag arcsec}^{-2}$ is approximately
uniform in an area of $300 \hbox{ arcmin}^2$, 
we find $L_{tot} \approx 3 \times 10^8 L_{\odot}$.
This value will be a lower limit if the fan extends further at fainter 
magnitudes or if there is more diffuse light in the inner regions.

The surface brightness in $M_{\odot}/pc^2$ 
of a simulated fan is calculated along a slit
with length $l=300$ kpc laid along the $x$ axis.  Particles are distributed
onto a Cartesian grid with cell size $\Delta l=3.75$ kpc in the $x - z$ 
projection.  To calculate
the surface brightness in $\hbox{mag arcsec}^{-2}$ in the Johnson B band,
a mass-to-light ratio must be assumed in order to transform 
$\Sigma(M_{\odot}/pc^2)$ to $\Sigma(L_{\odot}/pc^2)$.  
Then $\mu_B=24.3 - 2.5 log(\Sigma)$.  In Figure 13,
the results for Runs H1, H2, H6, and H9 are shown at two different times.
The length of the ordinate axis is 10 magnitudes as shown in the top left 
frame.
The four horizontal lines show $\mu_B=28 \hbox{mag arcsec}^{-2}$ for 
$M/L=$ 1, 2, 5, and 10 in ascending order as marked on the top right frame.
The limiting value for $\mu_B$ from the M87
observation is $28 \hbox{mag arcsec}^{-2}$ at approximately 110 kpc along the
$-x$ axis.

Models with higher-mass intruders than these will not satisfy the
observational constraints on the fan's surface brightness; 
too much material will be distributed 
along the $-x$ axis to account for the cut-off at $28 \hbox{mag arcsec}^{-2}$ 
at 100 kpc scales unless the mass-to-light ratio is
much larger than $M/L=10$.
This is true, for instance, of Run H7.  On the other hand, 
models like Run H5, with lower-mass intruders, do not create
a fan similar to that of M87.
The material is not dispersed enough to match
the observed distribution of diffuse material.

The last time (dotted line) shown for Run H1 is 
allowed with $M/L=10$  because the surface brightness decreases below
$28 \hbox{mag arcsec}^{-2}$ near 100 kpc, as are the times shown for Run H2.  
Earlier times, with more mass in the fan, are not
acceptable with unless the mass-to-light ratio is larger.  
Both times shown for Run H6 are acceptable for $M/L=$ 10 and 5.
And both times
shown for Run H9 are acceptable for $M/L=$ 10 and, possibly, 5.
None of these models fit the observational constraints best with a 
mass-to-light ratio of unity; all must be considered to contain some amount of
dark matter.
The surface brightness of most model fans 
has fallen below the observed limit in the Johnson B band by
$\approx 7\times 10^8$ years.
Diffuse fans will appear to be transient phenomena at this limit, 
persisting for only a few times $10^8$ years.  
Most of the debris will continue to stream into the central regions of
the primary galaxy.
Some unbound material, however,
will disperse into the region surrounding the primary galaxy.

The total luminosity in model fans is determined by
calculating the total mass over the 
area covered by the observed fan.  The mass-to-light ratio
which is allowed by the observed limiting value of $\mu_B$ is applied
to transform $M_{\odot}$ to $L_{\odot}$.  
For Run H1 at time 260, 
the total luminosity in an area of the fan similar to that of M87 is 
approximately $L_{tot}=4 \times 10^8 L_{\odot}$ with $M/L=10$.
For Run H2 at time 260, it is $L_{tot}=2 \times 10^8 L_{\odot}$ with $M/L=10$. 
With $M/L=10$, for Run H6 at time 260,  
$L_{tot}= 2.1 \times 10^8 L_{\odot}$ and, at time 280, 
$L_{tot}= 1.3  \times 10^8 L_{\odot}$.  
With $M/L=10$, for Run H9 at time 210, 
$L_{tot}= 2.6 \times 10^8 L_{\odot}$ and, at time 230, 
$L_{tot}= 1.8 \times 10^8 L_{\odot}$. 
The luminosity is twice these amounts if $M/L=5$. 
Given the uncertainty in
the calculation of total luminosity for the M87 fan, all of these values
are acceptable in comparison, although they imply that the mass-to-light
ratio of the intruder galaxy is likely close to or larger than $M/L=5$.

Kinematic anomalies suggest that 
subcomponents of accreted material may exist in M87.  
We calculated projected velocity fields for
intruder material along slits with length $l=60$ kpc
laid parallel to the indicated axis 
for Run H6 at time 260 and Run H9 at time 210.
The frames of Figure 14 show 
mean velocity $v_r$ (left) and velocity dispersion $\sigma$ (right) for 
intruder debris projected onto an intrinsic plane.
The observable line of sight velocities are in the $x-z$ projection
(the vantage point from which the fans are similar to that of M87).

The Run H6 and Run H9 $v_r$ results in the $x-z$ projection exhibit
deviations from zero, on the order of $50 - 150 \hbox{km s}^{-1}$.
Interestingly, this is similar to the velocity difference of the
kinematically distinct subcomponent seen in M87 by Jarvis \& Peletier
(1991), but, if observable, the extent of the disturbance is several
times larger in the models than is the blue-shifted annulus they see.
The debris which resides in the center of the models also has a very
large line-of-sight velocity dispersion along the x-slit.  The
projected dispersions are more than twice as large as those observed
in the center of M87.  There is little deviation from a smoothly
declining profile, and no departures from the profile that could
account for those observed using a long-slit spectrograph by Sembach
\& Tonry (1996).  Although it would be serendipitous to exactly
reproduce the M87 velocity profiles and although particle number and
cell width limit the resolution, these results do demonstrate that
mergers which produce diffuse fans may also produce small kinematic
disturbances.  However, the very large velocity dispersion of the intruder 
debris will make its influence on the kinematics of the integrated light
difficult to detect.  Broad shallow wings are easy to miss.

\section{Discussion}

The discovery of a diffuse fan in M87 motivated this study of 
the formation of stellar
debris at 100 kpc scales through large-mass-ratio accretion.  
The production of diffuse fans is not constrained by the shape of the 
primary potential, except in the sense that a very large core prevents
their formation.
The large core in the logarithmic model prevents the dense central mass
concentrations necessary to produce the accelerations that lead 
to large, diffuse fans of intruder debris.  On the other hand, the cuspy 
Hernquist potential and the higher-density cores of the Plummer and perfect
potentials easily form fans with large opening angles.
Fans formed during accretion into highly-flattened 
models are too
narrow to resemble the diffuse material in M87; the thin debris is interesting 
in the context of other of the Deep UK Schmidt Survey galaxies.
However, our results do
not disallow oblate, prolate, or spherical potentials for M87.

The debris in a spherical Hernquist potential has opening angles large
enough to easily account for the distribution of material in the
observed M87 fan.  Figure 15 shows a composite of M87 and Run H1.
Plummer and primary potentials produce
comparable but narrower distributions.  Although there is no evidence
to suggest that the mass density axis ratios in M87 are flattened, as
are the photometric axis ratios, the more flattened Hernquist $c/a=0.8$
models produce debris that resembles the M87 diffuse fan.  This is
consistent with studies of polar rings around elliptical and S0
galaxies such as NGC 4650A, where polar ring kinematics
indicate that the halo has flattened isodensity surfaces with axis
ratio $0.3 \leq c/a \leq 0.4$ (Sackett \etal 1994).  
In addition, simulations of merging pairs
or groups of spiral galaxies $-$ each with a disk, sometimes a bulge,
and a spherical dark-matter halo $-$ produce elliptical-like remnants
with flattened halos with $c/a \approx 0.8$ for the pair merger
remnants and $c/a \approx 0.8 - 0.9$ for multiple merger remnants
(Hernquist 1992, 1993, Weil \& Hernquist 1996).

The appearance of the M87 diffuse material does limit the orbital parameters
of the merging galaxy; large impact parameter orbits are ruled out.
In plunging orbits close to the major axis, the intruder passes near the 
center of the primary and is disrupted, its debris being spread out along the
major axis.  
If M87 is nearly stationary in the Virgo cluster core, 
high-velocity accretion
events like these could be common.  M87 would cannibalize galaxies on radial 
orbits with respect to the center of the cluster potential.
In order to utilize more than morphological information, the characteristics 
of the intruder debris in likely models are compared to
the limiting magnitude of the fan material and observations of kinematic 
substructure in M87.  Large projected velocity dispersions 
are seen in the $x-z$ plane. 
Although our resolution does not reach to the
100 pc range as do subarcsec resolution spectra (Jarvis \& Peletier 1991, 
van der Marel 1994), mean velocity can vary by 
$50 - 150 \hbox{ km s}^{-1}$ in the $x-z$ slits near the center.  
While much of the starlight in the M87 fan is too faint to be
accessible to kinematic analysis, embedded planetary nebulae are highly 
effective tracers (\eg, Hui \etal 1995).  Planetary
nebulae kinematics may allow confirmation of our model out to large radial
scales.
If these small accretions are common, however, 
there is no reason to expect that
the merger which produced the fan is also responsible for the central
$v_r$ variations.  

As the disrupted intruder evolves, 
debris disperses into the region surrounding the primary galaxy.  
At late times some of the intruder mass remains in low surface brightness 
($\mu_B \simlt 32 \hbox{mag arcsec}^{-2}$)
features several hundred kpc from the center of the galaxy.  This material
cannot easily account for a large amount of either halo or
intracluster medium mass.  As an upper limit, if similar accretions 
releasing a few $\times 10^9 M_{\odot}$ occur every $10^8$ years, only a few 
$\times 10^{11} M_{\odot}$ can be added to the halo or ICM.  
In addition, because most of the intruder mass remains bound to the 
primary galaxy, it is unlikely that these encounters can account for the
diffuse stellar light seen in clusters (\eg Vilchez-Gomez \etal 1994) except
in regions within 100 kpc of a large galaxy.
While this does not 
preclude more violent encounters in which most of the intruder is dispersed
widely, the diffuse matter can provide only a small fraction of the
unseen mass or intracluster light.  
In the same way, each intruder would be required to contribute a
hundred globular clusters to M87 at the rate of one encounter every 
$10^8$ years
to account for the observed excess.  Because encounters of the types we
model here do not leave behind compact remnants,
we could not expect to find multiple nuclei in the core of the primary.
Several accretion events, in which the intruder material has been dispersed 
below the threshold of present detection limits, could be hidden in
galaxies.  

M87 is not unique among massive elliptical galaxies in having a
dispersed fan of faint material along its major axis. 
The Deep UK Schmidt Survey of Nearby Galaxies has uncovered faint stellar
structure in NGC 4168 (also in Virgo), NGC 1316 (Fornax A),
NGC 2855 and NGC 5266.  Several of these ellipticals 
reveal other evidence for
accretion, including dust lanes, extended H\,I emission, and shells.
The spiral galaxies M104 and NGC 4643 are also embedded in a faint halo with
diffuse structure.  These observations suggest that the dispersal of
low-mass intruders over a large area is common 
(Malin, Weil, \& Bland-Hawthorn 1997).  This phenomenon has not been
recognized previously due to the very low surface brightness of the 
resulting distribution of material.  
The debris produced in systems resembling M87
generate kinematic substructure, in addition to low surface brightness, diffuse
fans.  We expect that future deep studies will continue to reveal diffuse
distributions of intruder debris and deviations to kinematic profiles on 
extended scales.  Although spectroscopy below $B = 24 \hbox{mag arcsec}^{-2}$
is difficult, one method of uncovering the substructure is to model the
underlying smooth elliptical out to the limits of the data and examine the
residuals.  Many studies have shown that numerous ellipticals have anomalies
in their rotation curves (Jedrzejewski \& Schechter 1989,
Franx, Illingworth, \& de Zeeuw 1991, Fried \& Illingworth 1994).  
Minor axis rotation has been used as a test for triaxiality (Binney 1985).
Anomalous changes in major axis rotation curves has been attributed to the 
capture of a small, compact companion (Jedrzejewski \& Schechter 1988).
Our observations and simulation results suggest that kinematic
disturbance in many ellipticals may be contamination 
by accreted debris, making
probes for higher order structure difficult.

\acknowledgments
MLW acknowledges funding by a PPARC postdoctoral fellowship.  
JBH is indebted to Oxford University for a Visiting Fellowship during the
summer of 1996.
We thank an anonymous referee for pointing out an important early reference
which had been missed previously and Tad Pryor for insightful and helpful
comments which greatly improved the manuscript.

\clearpage

\figcaption{\label{fig1} Image of M87 from photographic stacking of
five deep, IIIa-J (395-530nm passband) UK Schmidt plates.  The
limiting (B) surface brightness is $28 \hbox{mag arcsec}^{-2}$.}

\figcaption{\label{fig2} Digital Sky Survey image of M87, with length
$7^{\prime}$ and scaling mode set to wrapped linear.}

\figcaption{\label{fig3} Density profiles for several potentials along $x=y=z$.
Thin and thick lines are for a core or scale-length of 6 kpc and 25 kpc,
respectively.}

\figcaption{\label{fig4} Three times during the evolution of three logarithmic
models with masses $M_p = 1 - 3 \times 10^{13} M_{\odot}$.  Top panels show a 
spherical primary with core radius $R_c=6.0$ kpc, Run L1; middle panels show a
flattened primary with $q=0.85$ and $R_c=6.0$ kpc, Run L2; bottom panels show a
flattened primary with $q=0.85$ and $R_c=2.0$ kpc.  
The intruders are Plummer models which 
initially have mass $M_s = 1.2 \times 10^{10} M_{\odot}$, scale length
$a_s=1200$ pc, and initial positions $x_i=119$ kpc, $z_i=-13.7$ kpc. 
The length of each panel is 360 kpc and times in 
simulation units, one of which equals $1.2 \times 10^6$ years,
are shown at the top of each.}

\figcaption{\label{fig5} 
Snapshots of intruder debris for three spherical primaries: Hernquist 
(top, Run H1),
Plummer (middle, Run P1), and perfect (bottom, Run P2).  Each has 
$M_p= 3 \times 10^{13} M_{\odot}$, $a_p=6$ kpc, and a Plummer model 
intruder with 
$M_s = 1.2 \times 10^{10} M_{\odot}$, scale length $a_s=1.2$ kpc, 
and initial positions $x_i=119$ kpc, $z_i=-13.7$ kpc.
}

\figcaption{\label{fig6} 
Time evolution for Run H2, an oblate Hernquist potential with $c/a=0.8$ and
the fiducial intruder of Figure 5. The
length of each panel is 360 kpc and times in simulation units are
shown at the top of each panel.  The ellipses show the shape of the
primary surface density at $x =$ 50 and 120 kpc.  The top row shows
initial position in the $x - z$, $x - y$, and $y - z$ projections.
Subsequent panels show the evolution in the $x - z$ plane.}

\figcaption{\label{fig7} Snapshots of intruder debris for an oblate Hernquist 
primary with $c/a=0.6$ (Run H3) and a prolate Hernquist primary with $c/a=0.8$
(Run H4).}

\figcaption{\label{fig8} Snapshots of intruder debris for three 
oblate Hernquist primaries with $c/a=0.8$.  In the top panels, the intruder
has $M_s = 1.2 \times 10^{9} M_{\odot}$ and $a_s=300$ pc (Run H5).
In the middle panels, the intruder has $M_s= 6.0 \times 10^{9} M_{\odot}$
and $a_s=600$ pc (Run H6).  In the bottom panels, the intruder has
$M_s= 6.0\times 10^{10} M_{\odot}$ and $a_s=2.0$ kpc (Run H7).}

\figcaption{\label{fig9} Velocity  histogram for Runs H5, H2, and H7,
respectively.  The top frames are for times at which the particles are 
bound to the secondary potential; the bottom frames are for 
subsequent times, when the most of the mass has just become unbound.}

\figcaption{\label{fig10} Time evolution for Run H8, an oblate Hernquist 
potential with $c/a=0.8$ but with $a_p=25$ kpc.}

\figcaption{\label{fig11} Snapshots of intruder debris for primaries with 
$M_p = 5 \times 10^{12} M_{\odot}$ and the fiducial intruder.  
Top panels are for Run L3 with a logarithmic primary.  Middle panels are
for Run H9, with an oblate Hernquist primary with $c/a=0.8$.  Bottom
panels are for Run H10, similar to Run H9 except that $x_i=59.4$ kpc and
$z_i=-8.5$}
 
\figcaption{\label{fig12} Evolution of Runs H1 (top), H2 (middle), and H5 
(bottom) at lates times.}

\figcaption{\label{fig13} Projected surface brightness profile in
Johnson B-band magnitudes for a slit with length $l=300$ kpc laid
along the x axis.  The two different times for Runs H1, H2, H6, and H9 are
shown chronologically with a dashed and dotted line in each
frame.  The ordinate axis is 10 magnitudes in length and the effect of
varying $M/L$ is represented by the four horizontal lines which show
$\mu_B=28 \hbox{mag arcsec}^{-2}$ for $M/L=$ 1, 2, 5, and 10 as
detailed in the second frame.}

\figcaption{\label{fig14} Projected velocity fields for Run H6 at times 
260 and Run H9 at time 130.
Slits of length $l=60$ kpc 
are laid parallel to the indicated axis.  Frames show mean velocity 
$v_r$ (left) and velocity dispersion $\sigma$ (right) for
projections onto the $x-z$ planes.}

\figcaption{\label{fig15} M87 overlaid by $x-z$ projection of Run H1
at time 260.  Ellipses show the shape of the M87 model surface density
at 50 and 120 kpc.}

\clearpage

\begin{table}[b]
\centering
\small
\caption[Subset of Companions]
{\label{tab:m87t1}Subset of Systems}
\begin{tabular}{lllllll}
\multicolumn{7}{c}{} \\ \hline
\multicolumn{1}{c}{Run}&\multicolumn{1}{c}{$M_p (M_{\odot})$}&
\multicolumn{1}{c}{$M_s (M_{\odot})$}&
\multicolumn{1}{c}{$a_s$ (kpc)}&
\multicolumn{1}{c}{$x_i$ (kpc)}&\multicolumn{1}{c}{$z_i$ (kpc)}&
\multicolumn{1}{c}{Primary}\\ \hline
\multicolumn{7}{c}{} \\
L1&$2.4\times 10^{13}$&$1.2\times 10^{10}$&1.2&119&-13.7&Logarithmic q=1.0\\
L2&$3\times 10^{13}$&$1.2\times 10^{10}$&1.2&119&-13.7&Logarithmic q=0.85\\
L3&$10^{13}$&$1.2\times 10^{10}$&1.2&119&-13.7&Logarithmic q=0.85\\
H1&$3\times 10^{13}$&$1.2\times 10^{10}$&1.2&119&-13.7&Hernquist spherical\\
H2&$3\times 10^{13}$&$1.2\times 10^{10}$&1.2&119&-13.7&Hernquist c/a=0.8\\
H3&$3\times 10^{13}$&$1.2\times 10^{10}$&1.2&119&-13.7&Hernquist c/a=0.6\\
H4&$3\times 10^{13}$&$1.2\times 10^{10}$&1.2&13.7&119&Hernquist prolate\\
H5&$3\times 10^{13}$&$1.2\times 10^{9}$&0.3&119&-13.7&Hernquist c/a=0.8\\
H6&$3\times 10^{13}$&$6.0\times 10^{9}$&0.6&119&-13.7&Hernquist c/a=0.8\\
H7&$3\times 10^{13}$&$6.0\times 10^{10}$&2.0&119&-13.7&Hernquist c/a=0.8\\
H8&$3\times 10^{13}$&$1.2\times 10^{10}$&1.2&119.5&-11.0&Run H2 with $a_p=25$\\
H9&$5\times 10^{12}$&$10^{10}$&1.2&119&-13.7&Hernquist c/a=0.8\\
H10&$5\times 10^{12}$&$10^{10}$&1.2&59.4&-8.5&Hernquist c/a=0.8\\
P1&$3\times 10^{13}$&$1.2\times 10^{10}$&1.2&119&-13.7&Plummer spherical\\
P2&$3\times 10^{13}$&$1.2\times 10^{10}$&1.2&119&-13.7&perfect spherical\\
\multicolumn{7}{c}{} \\ \hline
\end{tabular}
\end{table}

\end {document}